\documentclass{ifacconf}

\usepackage{graphicx}      
\usepackage{natbib}        

\usepackage{graphicx}      
\graphicspath{{imgs/}}
\usepackage{natbib}        
\usepackage{soul, color}
\usepackage{amsmath}
\usepackage{amsfonts}
\usepackage{tikz}
\usepackage{enumitem}

\renewcommand{\mod}{\mathrm{mod}}

\usepackage{xfrac}

\newtheorem{Definition}{Definition}[section]
\newtheorem{Assumption}{\textbf{Assumption}}[section]
\newtheorem{Remark}{\textbf{Remark}}[section]
\newtheorem{Procedure}{Procedure}

\usepackage{color}
\definecolor{mblue}{rgb}{0,0.447,0.741}
\definecolor{mred}{rgb}{0.85,0.325,0.098}
\definecolor{myellow}{rgb}{0.9290,0.6940,0.1250}

\begin{document}
\begin{frontmatter}

In\textit{ 21st IFAC World Congress}, Berlin, Germany, 2020 
\title{Gaussian Process Repetitive Control for Suppressing Spatial Disturbances} 

\thanks[footnoteinfo]{This research received funding from the European Union H2020 program under grant agreement 637095 (FourByThree), and ECSEL-2016-1 under grant agreement 737453 (I-MECH), and VIDI with project number 15698 (partly) financed by the Netherlands Organisation for Scientific Research (NWO).}

\author{Noud Mooren$^\mathrm{*}$  Gert Witvoet $^\mathrm{*, **}$ Tom Oomen $^\mathrm{*}$} 

\address[First]{Control Systems Technology, Department of Mechanical Engineering, Eindhoven University of Technology, Eindhoven, The Netherlands (n.f.m.mooren@tue.nl, g.witvoet@tue.nl, t.a.e.oomen@tue.nl)}
\address[Second]{TNO Technical Sciences, Optomechatronics Department, Delft, The Netherlands.}

\begin{abstract}                
Motion systems are often subject to disturbances such as cogging, commutation errors, and imbalances, that vary with velocity and appear periodic in time for constant operating velocities. The aim of this paper is to develop a repetitive controller (RC) for disturbances that are not periodic in the time domain, yet occur due to an identical position-domain disturbance. A new spatial RC framework is developed, allowing to attenuate disturbances that are periodic in the position domain but manifest a-periodic in the time domain. A Gaussian process (GP) based memory is employed with a suitable periodic kernel that can effectively deal with the intermittent observations inherent to the position domain. A mechatronic example confirms the potential of the method.

\end{abstract}


\end{frontmatter}

\section{Introduction}
Motion systems performing repeating tasks often generate disturbances that are periodic in the position domain. Consider for example; a rotational system with bearings or imperfections leading to an imbalance, a motor/gear system where a tooth profile induces a disturbance, or a linear motor with cogging \citep{AhnCheDou2005}. These are typical examples where disturbances are inherently spatially periodic, i.e., depending on (angular-) position. For constant operating conditions, i.e., a constant angular velocity or a system performing repeating tasks, these disturbances appear periodic in the time domain. However, if the operating conditions vary over time, the disturbance will appear a-periodic in the time domain, see e.g., \cite{ChenChi2008}, \cite{Li2015}.

Repetitive control (RC) attenuates periodic disturbances with a fixed and known period, see e.g., \cite{HaraYamOmaNak1988, Longman2010, WangGaoDoy2009}. For constant operating conditions, position-domain disturbances manifest periodically in the time domain. Hence, only then RC can suppress the disturbance by capturing an internal model in a time-domain memory loop. Subsequently, by the internal model principle \citep{FrancisWonham1976} RC can asymptotically reject the periodic disturbance. 

Traditional RC approaches lead to a degraded performance if the operating velocity, i.e., disturbance frequency, changes. Many practical applications require a continuously varying operation velocity, for example a printer moving over paper, a rotary system for tracking satellites \citep{Saathof2019}, or a wafer-scanner \citep{BlankenBoeBruOom2017}. In these cases, the disturbance is periodic in the position domain, however, due to speed variations it appears a-periodic in the time domain. This results in a situation where traditional RC is not effective.

Several repetitive control techniques have been developed to deal with these period variations. In \cite{BlankenOom2019}, an extension is presented allowing to design multiple RCs for systems that operate at multiple fixed velocities. However, performance during velocity changes is not guaranteed. In \cite{WitvoetPetKuiOom2019}, an alternative is presented where the fundamental disturbance frequency is chosen as small as possible covering a wide range of disturbance frequencies, which is rather pragmatic and may lead to worse stability margins. In \cite{Steinbuch2002}, RC has been extended with robustness against small variations in the disturbance frequency, however large or continuous variations are not covered. Other existing approaches focus on defining the system in the spatial domain, resulting in a nonlinear system that requires an additional step of feedback linearization, see e.g., \cite{ChenYang2007}.

Although recent progress has been made to increase robustness for slightly varying disturbance frequencies, a solution for large or continuous variations, where the source of the disturbance is repeating in the position domain, is not yet established. In answer to this, the aim of this paper is to extend RC towards the spatial domain. The key idea is to use a spatial memory, instead of a time-domain memory, by means of a Gaussian Process (GP) with a suitable periodic kernel \citep{Murphy2012, Rasmussen2006, PillonettoDinCheNicLju2014, JidlingHenWahGreSchoWenWill2018}. The GP estimates a continuous function, allowing to deal effectively with the intermittent observations as they occur in the position domain. This paper contains the following sub-contributions;
	\begin{enumerate}[label=(C\arabic*)]
		\item spatial RC approach for systems with a-periodic and varying disturbance frequencies,
		\item a new spatial memory with a GP and suitable periodic kernel allowing for intermittent observations, in combination with a time domain learning filter, and
		\item a simulation case study confirming the potential. 
\end{enumerate}

This paper is outlined as follows. The disturbance rejection problem is defined in section \ref{sec:problem_definition}. In Section \ref{sec:PB_RC_setting}, the spatial RC approach is introduced (C1). In Section \ref{sec:BP_buffer_GP}, a position-domain memory loop by means of a GP is provided with a suitable periodic kernel (C2). In Section \ref{sec:Simulation_example}, a simulation example shows that spatial RC outperforms the standard RC in the considered setting (C3). Finally, conclusions are provided in Section \ref{sec:conclusions}.


\section{Problem Definition} 
\label{sec:problem_definition}
In this section, the problem setting and considered disturbances are defined. This shows that observations indeed are non-equidistant in position. Finally, the problem definition and contributions are formulated.

\begin{figure}[b]
	\begin{centering}
		\includegraphics[page=10, width=.8\linewidth]{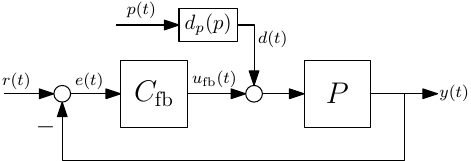}
		\caption{Problem setting.}
		\label{fig:problem_setup}
	\end{centering}
\end{figure}

\subsection{Problem setting}
\begin{figure}
	\begin{centering}
		\includegraphics[width=0.9\linewidth]{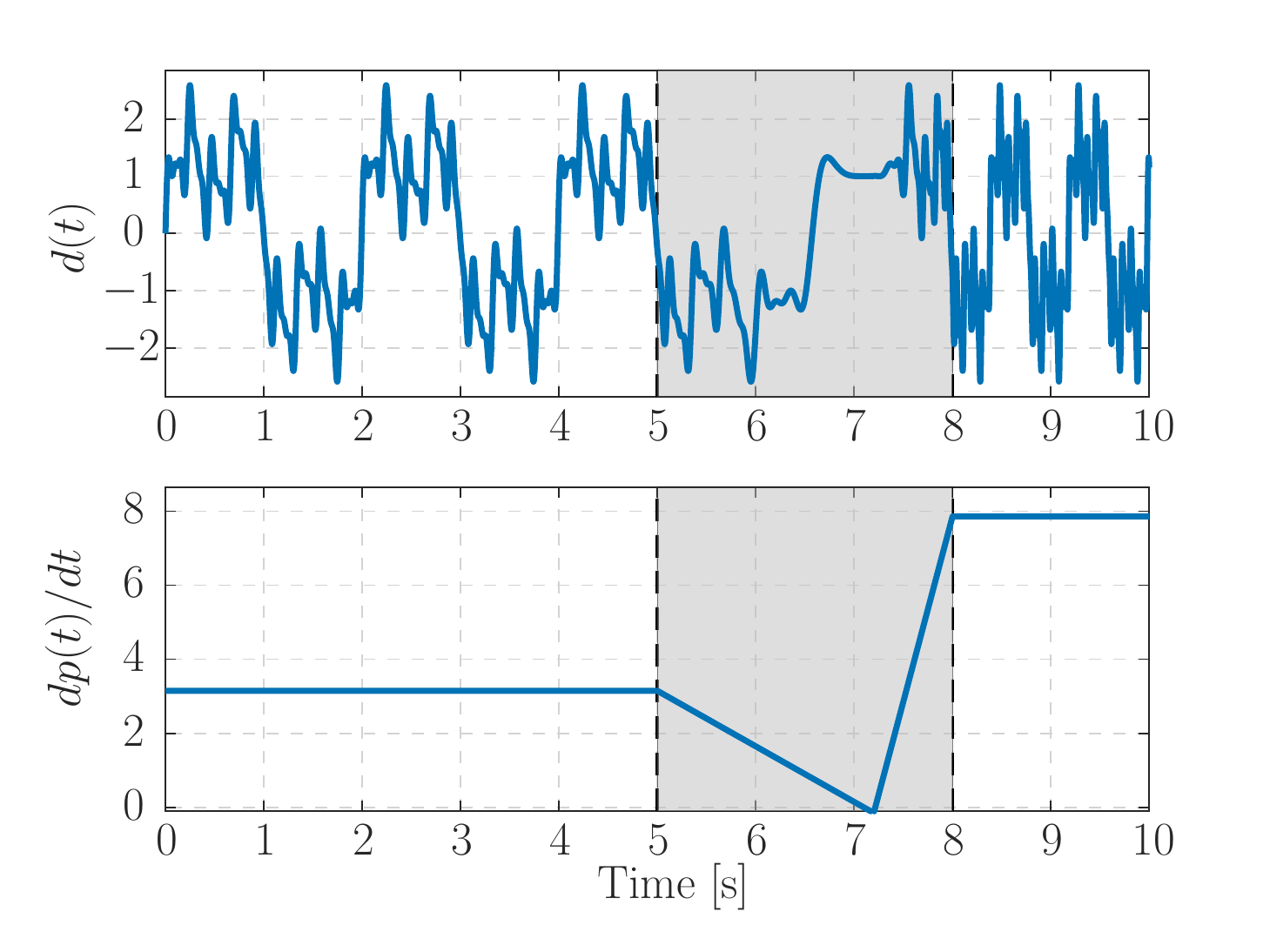}
		\caption{Simulation example of the disturbance (top plot), with corresponding velocity (bottom plot) as function of time. Gray area indicates the velocity change causing a-periodicity in the time domain disturbance.}
		\label{fig:Dist_var_velocity}
	\end{centering}
\end{figure}

Consider the control setup depicted in Fig. \ref{fig:problem_setup}, where $C_\mathrm{fb}$ is a stabilizing feedback controller, and $P$ is a single-input single-output (SISO) linear time-invariant (LTI) plant. The signal $r(t)$ is an optional reference to be tracked, $y(t)$ is the output to be controlled, and $d(t)$ is an unknown exogenous disturbance that is collocated with the input signal $u_\mathrm{fb}(t)$. The disturbance $d(t)$ is assumed
to be periodic in the spatial domain with some given spatial period $p_\mathrm{per}$, i.e., it satisfies the following assumptions.
\begin{Assumption} \emph{Let $d(t)$ be an unknown exogenous disturbance that satisfies: 
	\begin{enumerate}[label=(A\arabic*)]
		\item $d(t)$ is composed of an unknown position-domain function $d_p(p)$ and the current position $p(t)$, i.e.,
		\begin{align}
		d(t) = d_p(p(t))
		\label{eq:dist_gen}
		\end{align}
		which is schematically depicted in Fig. \ref{fig:problem_setup}, and
		\item $d_p(p)$ is spatially periodic with period $p_\mathrm{per}$, i.e., \begin{align} d_p(p) = d_p(p+k\cdot p_\mathrm{per})\; \text{for} \; k \in \mathbb{N} \end{align} where $p_\mathrm{per} \in \mathbb{R}$ is known and fixed.
	\end{enumerate}}
	\label{ass:disturbance}
\end{Assumption}
The disturbance $d(t)$ depends on the position signal $p(t)$. If the position increases/decreases with a fixed rate, i.e., the speed $\sfrac{d p(t)}{d t}$ is constant, then $d(t)$ appears periodic in the time domain. If the velocity is time varying, then the disturbance is a-periodic in the time domain, while the underlying function generating the disturbance $d_p$ remains periodic in the position domain. An example is shown in Fig. \ref{fig:Dist_var_velocity}, where the velocity changes in the gray area leading to a-periodicity in the time domain.

The aim of this paper is to develop a spatial RC approach such that the following control objective is satisfied.
\begin{Definition}[Control Objective]
\emph{Minimize the effect of the potentially a-periodic disturbance $d(t)$ on the position error $e(t)$, independent of velocity variations.}
\end{Definition}

\subsection{Traditional repetitive control}
Traditional repetitive control is not effective for disturbances with varying period or a-periodic disturbances, see, e.g., \cite{Longman2010, HaraYamOmaNak1988}. According to the well-known internal model principle, a model of the disturbance must be present in the feedback loop to attenuate it \citep{FrancisWonham1976}. In traditional RC, this is captured in a time-domain memory loop based on previous error data. The size of the memory loop must coincide with the disturbance period. If the disturbance period changes, the pre-defined buffer size is not adequate anymore, resulting in degraded performance. 

The key idea in this paper, is to specify a memory loop in the spatial domain, where the disturbance is assumed to be fixed and periodic. The main challenge arising in the spatial domain, is that observations are inherently non-equidistantly distributed due to speed variations. Hence, a fixed memory loop as in traditional RC is not suitable, see \cite{MoorenWitAcaKooOom2020}. Therefore, an alternative solution using a GP, essentially estimating a continuous memory loop, is presented here.

\subsection{Problem definition}
Disturbances that vary in the time domain, but yet have an underlying position-domain disturbance that is periodic, cannot be attenuation by traditional RC. This necessitates the construction of a memory loop in the position domain to deal with the non-equidistant nature of the observations. Thus, the aim of this paper is to develop a new spatial RC approach, in which a memory loop is constructed in the spatial domain by means of a Gaussian process with a suitable kernel. This enables suppression of disturbances satisfying (A1)-(A2) with varying frequency.

\section{Spatial Repetitive Control Framework}
\label{sec:PB_RC_setting}
In this section, the spatial RC framework is established, enabling attenuation of disturbances that are periodic in the spatial domain. The subsequent section is devoted to the of a spatial memory loop using a GP.

\subsection{Spatial repetitive control setting}
\begin{figure}
	\begin{centering}
		\includegraphics[page=5, width=\linewidth]{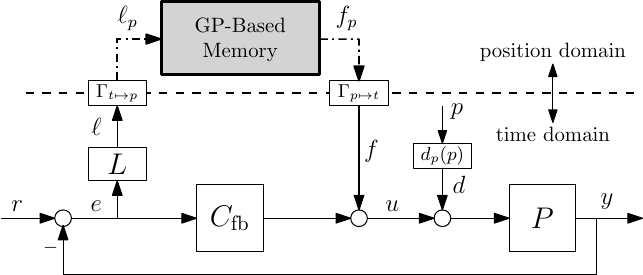}
		\caption{Position-based repetitive control scheme, where the solid lines represent time-domain signals and the dashed lines represent position-domain signals.}
		\label{fig:RC_setup_pos}
	\end{centering}
\end{figure}

The spatial RC framework is depicted in Fig. \ref{fig:RC_setup_pos}, where $L$ is a learning filter, $\ell(t)$ is the learning signal, and $f(t)$ is the output of the RC that is injected in the feedback loop. An explicit distinction is made between time-domain signals and position-domain signals with subscript $(\cdot)_p$. The learning signal $\ell(t)$ is mapped to the position-domain signal $\ell_p(p)$, indicated by the mapping $\Gamma_{t \mapsto p}$. Subsequently, the signal $\ell_p(p)$ is used to construct the disturbance model in the spatial memory loop. The output of the memory loop is mapped back to the time domain by $\Gamma_{p \mapsto t}$, resulting in $f(t)$ that is injected in the feedback loop. Note that the learning filter $L$ is located before the buffer, which will appear to be essential in this spatial approach. 

\begin{figure}
	\begin{centering}
		\includegraphics[page=6, width=0.8\linewidth]{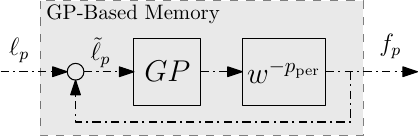}
		\caption{Spatial memory loop with a Gaussian process.}
		\label{fig:RC_setup_buffer}
	\end{centering}
\end{figure}

The generated signal $f(t)$ exactly compensates for the disturbance $d(t)$. This is done by estimating the function $d_p(p)$, i.e., the underlying cause of the disturbance, in the memory loop by means of a GP, see Fig. \ref{fig:RC_setup_buffer}. Details on the implementation of the GP in the memory loop are discussed in Section \ref{sec:BP_buffer_GP}. To estimate $d_p$ in the buffer, a specific choice of learning filter is required, which is analog to the $L$ filter design in traditional RC.
\vspace{4pt} \hrule \vspace{2pt}
\begin{Procedure}[Learning filter design] \hfill 
	\vspace{2pt} \hrule
\begin{enumerate}
	\item Identify a parametric model $\hat P$ of the system,
	\item compute an estimate of the process sensitivity 
	\begin{align}
	\widehat{PS} = \frac{\hat P}{1+C_\mathrm{fb} \hat P}\; ,
	\end{align}
	\item invert the process sensitivity estimate to obtain $L$
	\begin{align} 
	L = \widehat{PS}^{-1}.
	\label{eq:learning_filter}
	\end{align} 
\end{enumerate}
	\label{proc:L_fitler}
	\vspace{4pt} 	\hrule
\end{Procedure}

\begin{Remark}
	If $\widehat{PS}$ contains non-minimum phase zeros, leading to an unstable inverse, several techniques exist to compute stable but non-causal inverses, see e.g., \cite{Zundert2017_Journal, Tomizuka1987}.
\end{Remark}
Note that the learning filter is allowed to have finite preview, i.e., non-causality $L = z^{n_l} L_c$, where $n_l$ is the number of samples of preview. How to include this in the memory loop is shown in Section \ref{sec:BP_buffer_GP}. 

The learning signal is mapped to the spatial domain to generate $\ell_p$, used in the memory loop as shown in Fig. \ref{fig:RC_setup_buffer}. 
\vspace{4pt} \hrule \vspace{2pt}
\begin{Procedure}[Mapping $\ell(t)$ to $\ell_p(p)$] \hfill 
	\vspace{2pt} \hrule
	Given the current position $p(t)$ 
	\begin{enumerate}
		\item obtain $p^*(t) = \mod \left ( p(t), p_\mathrm{per} \right )$
		\item construct $ \ell_p(p) = \ell(p)\circ p^*(t)$
	\end{enumerate}
	\label{proc:p_to_t}
	\vspace{4pt} 	\hrule
\end{Procedure}
By selecting the learning filter $L$ as the inverse of the process sensitivity, the learning signal $\ell(t)$ becomes an approximation of the disturbance, i.e., $\hat d(t) = L e(t).$ Hence, $\ell_p(p)$ will approximate $d_p(p)$ that is stored in the GP. Note that a feedback loop is present in the memory loop. This allows the error and therefore also $\ell_p(p)$ to converge to zero while $\tilde \ell_p(p)$ remains equal to an estimate of $d_p(p)$. Furthermore, the output of the memory loop is delayed with exactly one period in the spatial domain similar to traditional RC, denoted with $w^{- p_\mathrm{per}}$.

Finally, the output of the memory loop $f_p(p)$ is mapped to the time domain which completes the RC framework.
\vspace{4pt} \hrule \vspace{2pt}
\begin{Procedure}[Mapping $f_p(p)$ to $f(t)$] \hfill 
	\vspace{2pt} \hrule
	Given the current position $p(t)$,
	\begin{enumerate}
		\item obtain $p^* = \mod \left ( p(t), p_\mathrm{per} \right )$,
		\item evaluate the posterior mean of the GP's density function at $p^*$ given observations,  denoted as $$u_\mathrm{ff}(t) =\mu_\mathrm{post}(p^*)$$ and further discussed in Section \ref{sec:BP_buffer_GP}.
	\end{enumerate}
	\label{proc:t_to_p}
	\vspace{3pt} 	\hrule
\end{Procedure}
Several interesting observations can be made with respect to traditional RC. First of all, the learning filter $L$ is placed before instead of after the memory loop. In the tradition RC, the $L$ filter can be placed before or after the memory loop using the commutative property of linear systems. Because of the transformations from time to position domain this property does not hold anymore. Secondly, the summation point is moved to the output of the feedback controller instead of being at the input. This allows to extrapolate the learned function $d_p(p)$ to the time-domain, i.e., generate the exact opposite of $d(t)$. Finally, the mappings between time and position domain can be interpreted similar to using basis function in iterative learning control (ILC), see e.g., \citet{WijdevenBos2010}. Allowing to learn the disturbance in the position domain, and extrapolate to the time domain for varying operating velocities. 

In the following section, the implementation of GPs in the memory loop is discussed.

\section{Spatial memory loop using GP\small{s}} \label{sec:BP_buffer_GP}
A Gaussian Process (GP) regression model can be interpreted as a distribution over functions. Using inference on the basis of training data, i.e., observations, in combination with priors, represented by a kernel, one can determine statistical properties of the underlying function, see e.g., \cite{Murphy2012, Rasmussen2006}. In this section the GP is analyzed and a suitable periodic kernel is presented. Finally, a procedure is provided to integrate a GP in the spatial memory as shown in Fig. \ref{fig:RC_setup_buffer}.

\subsection{GP based spatial memory loop}
The aim is to estimate a continuous function representing the true spatial disturbance $d_p(p)$, using observations $\ell_p(p)$ that are contaminated with noise, i.e.,
\begin{align}
\tilde \ell_p(p) = d_p(p) + \epsilon, \text{ with } \epsilon \sim \mathcal{N}(0, \sigma_n^2) 
\end{align}
where $\epsilon$ is independent identically distributed zero-mean Gaussian noise with variance $\sigma_n^2$. Furthermore, a suitable kernel choice allows to deal deal with noisy observations and include desired properties of the estimated signal, such as periodicity and smoothness.

Next, assume that $d_p$ and $\tilde \ell_p$ are random variables that have a joint Gaussian distribution denoted as follows,
\begin{align}
\begin{bmatrix} d_p \\ \tilde \ell_p  \end{bmatrix} \sim 
\mathcal{N} \left ( \begin{bmatrix} 0 \\ 0 \end{bmatrix}, 
\begin{bmatrix}
K + \sigma_n^2 I_N & K_* \\ K_*^\top & K_{**}
\end{bmatrix} \right )
\end{align}
where $K \in \mathbb{R}^{N \times N}$, $K_* \in \mathbb{R}^{N \times N^*}$ and $K_{**} \in \mathbb{R}^{N^* \times N^*}$ are kernels or covariance functions, with $N$ the number of observations and $N^*$ the number of test positions. 

\begin{Remark}
	Note that the mean of the distribution is assumed to be zero for the ease of notation. This can easily be extended for a non-zero mean \citep{Murphy2012}.
\end{Remark}
The Gaussian process is completely determined by its mean and covariance function. Given observations $\tilde \ell_p$, the posterior distribution becomes,
\begin{align}
\hat d_p | \tilde \ell_p \sim \mathcal{N} \left (\mu_\mathrm{post}, P_N^\mathrm{post} \right ),
\end{align}
with mean $\mu_\mathrm{post}$ and covariance $P_N$ matrix, 
\begin{align}
\mu_\mathrm{post} &= K_*^\top (K+\sigma_n^2 I_N)^{-1} \tilde \ell_p,  \\
P_N &= K_{**} - K_*^\top (K + \sigma_n^2 I_N)^{-1} K_*.
\end{align} Since there is only one test point $p$ at a time, the posterior mean can be computed efficiently as follows,
\begin{align}
\mu_\mathrm{post}(p) = \sum_{i = 1}^N \alpha_i \kappa(p_i, p)
\label{eq:GP_post_mean_single}
\end{align}
where $\kappa(p_i,p)$ is the kernel evaluated at training point $p_i$ and test point $p$, and
\begin{align}
\mathbf{\alpha} = (K+\sigma_n^2 I)^{-1} \tilde \ell_\mathrm{train} = \begin{bmatrix} \alpha_1 & \alpha_2 & \hdots & \alpha_N \end{bmatrix} 
\label{eq:GP_post_mean_kernel}
\end{align}
in which $\tilde \ell_\mathrm{train}$ contains the observations at position $p_i$. 
\begin{figure}
	\begin{centering}
		\includegraphics[width=0.7\linewidth]{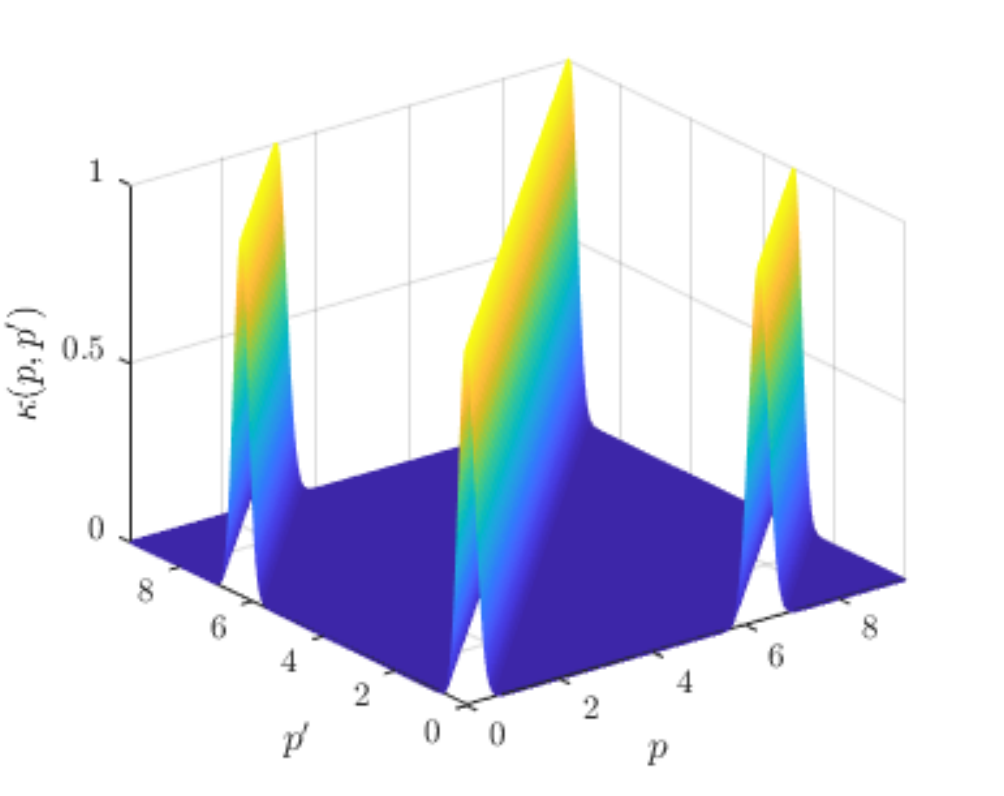}
		\caption{Periodic kernel used for spatial RC, with $\lambda = 2\pi$, $l = 0.2$, $\sigma_f = 1$ and $\sigma_n = 10^{-6}$.}		
		\label{fig:Kernel_periodic}
	\end{centering}
\end{figure}

\subsection{Periodic kernels}
Next, is the selection of a suitable kernel function imposing priors on the estimate $\hat d_p$. Since the disturbance $d_p$ is periodic this should be included in the kernel function. However, traditional kernels as often used in system identification approaches, see e.g., \cite{PillonettoDinCheNicLju2014, ChenOhlLju2012, BlankenOom2020_GP}, do not impose these type of priors. In this paper, the following kernel is utilized to impose periodicity and smoothness,
\begin{align} 
\kappa(p,p^\prime ) &= \sigma_f^2 \exp \left (\frac{-2 \sin^2(\frac{p-p^\prime}{2 \lambda})}{l^2} \right )
\end{align}
that is periodic due to the presence of the sine function, and $\sigma_f$, $\lambda$, $\ell$ are the hyper-parameters. An example of a periodic kernel for a specific set of hyper-parameters is depicted in Fig. \ref{fig:Kernel_periodic}, where it can be seen that the kernel repeats every $2 \pi$. The parameter $\lambda$ represents the period of the function $d_p$, which is given by $p_\mathrm{per}$ according to Assumption \ref{ass:disturbance}. This includes periodicity, i.e., an observation at $p = 0$ infers something about the position $p_\mathrm{per}$. The parameter $l$ imposes smoothness of the observations assuming that the underlying function $d_p$ is smooth as well. Hence, an observation at some position $p$ also provides information about positions close to that specific observation. This kernel allows to extrapolate beyond the data observations for fast learning. 

\begin{Remark}
	An empirical Bayesian optimization can be used to optimize the hyper-parameters, see e.g., \cite{SnoekLarAda2012}.
\end{Remark}

\subsection{GP buffer procedure}
In this section, a procedure to integrate GPs in the spatial RC framework is presented. Furthermore, non-causality of the learning filter is included in the memory loop.

\textbf{\hrule}
\begin{Procedure}[Position-domain RC using GP] \hfill 
	\hrule
	\textbf{I. Initialization and prior}
	\begin{enumerate}[leftmargin=1cm]
		\item Set the kernel parameters $\sigma_f$, $\sigma_n$, $\lambda$ and $l$.
		\item Obtain observation $\tilde \ell_p(p)$ at current position $p(k)$.
		\item Set counter $i = 1$
	\end{enumerate}
	\textbf{II. Every $\bar N^\mathrm{th}$ sample:}
	
	\hspace{0.3cm} \textbf{if $ p(k) \leq p_\mathrm{per}$}
	\begin{enumerate}[leftmargin=1.2cm]

		\item Store training observations: 
		\begin{align*}
		p_\mathrm{train}(i) = \tilde p, \quad \ell_\mathrm{train}(i) = \tilde \ell_p
		\end{align*}
		where $\tilde \ell_p = \ell_p$, and 
		\begin{align}
			\tilde p = \mathrm{mod}(p(k) - n_l\frac{d p}{dt}  T_s, p_\mathrm{per})
		\label{eq:p_preview}
		\end{align}
		with $n_l$ is the number of preview samples in $L$ and $T_s$ is the sample time.
		\item Set $i = i + 1$
	\end{enumerate}
		\hspace{0.3cm} \textbf{else}
	\begin{enumerate}[leftmargin=1.2cm]
		\item In this case the memory loop is already filled, which is included in the computation of $\tilde \ell_p$. Again store the data,
		\begin{align*}
		p_\mathrm{train}(i) = \tilde p, \quad \ell_\mathrm{train}(i) = \tilde \ell_p
		\end{align*}
		where $\tilde p$ is given by \eqref{eq:p_preview}, and $\tilde \ell_p = \tilde \ell + \mu_\text{prev}(\tilde p(k))$ and $\mu_\text{prev}$ is the posterior mean obtained by evaluation the GP given the data $p_\mathrm{train}$ and $\ell_\mathrm{train}$ corresponding to the previous spatial period.
		\item Set $i = i+1$		
	\end{enumerate}
	\textbf{III. Computing $u_\mathrm{ff}(k)$, at every sample}
	\begin{enumerate}[leftmargin=1.2cm]
		\item Given data $p_\mathrm{train}$ and $\ell_\mathrm{train}$ from the previous period, compute the posterior mean of the GP using \eqref{eq:GP_post_mean_single} and \eqref{eq:GP_post_mean_kernel}.
	\end{enumerate}
	\label{proc:RC_GP_batch}
	\hrule
\end{Procedure}

Note that every $\bar N^\mathrm{th}$ observation is used to update the GP. This facilitates the computational aspects and can easily be implemented in the kernel by adapting the smoothness parameter $l$. By setting $\bar N = 1$ all observation are used.

This completes the spatial RC framework including a GP based memory loop. In the following section, spatial and traditional RC are compared in a simulation example.
\begin{figure}[b]
	\begin{centering}
		\includegraphics[width=0.8\linewidth]{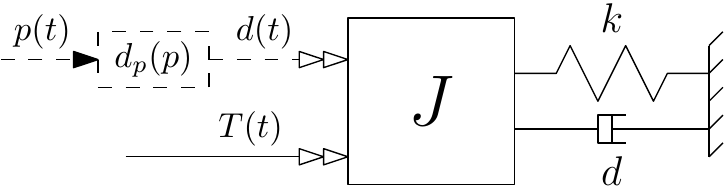}
		\caption{Simulation model with mass-spring-damper system, disturbance $d(t)$ and input torque $T(t)$.}
		\label{fig:MSD_dist}
	\end{centering}
\end{figure}

\section{Simulation study}  \label{sec:Simulation_example}
In this section, a simulation example is provided to confirm that the spatial RC approach outperforms traditional RC for changing disturbance frequencies.
\subsection{System description and simulation setting}
Consider a second order mass-spring-damper system,
\begin{align}
P(s) = \frac{1}{Js^2 +d s+ k}
\end{align}
with inertia $J = 1$ kg$\cdot$m$^2$, damping $d = 1$ Nm/s, and stiffness $k = 10^4$ Nm. A position-dependent disturbance $d(t)$ acts on the system being controlled by a torque $T$, see Fig. \ref{fig:MSD_dist}. The plant is discretized by zero-order-hold discretization with sample frequency $f_s = 1000$ Hz. A stabilizing feedback controller is designed consisting of a gain, lead filter and a low-pass filter. The resulting bandwidth, i.e., $0$ dB crossing of the open-loop, is $50$ Hz.
\begin{figure}
	\begin{centering}
		\includegraphics[width=\linewidth]{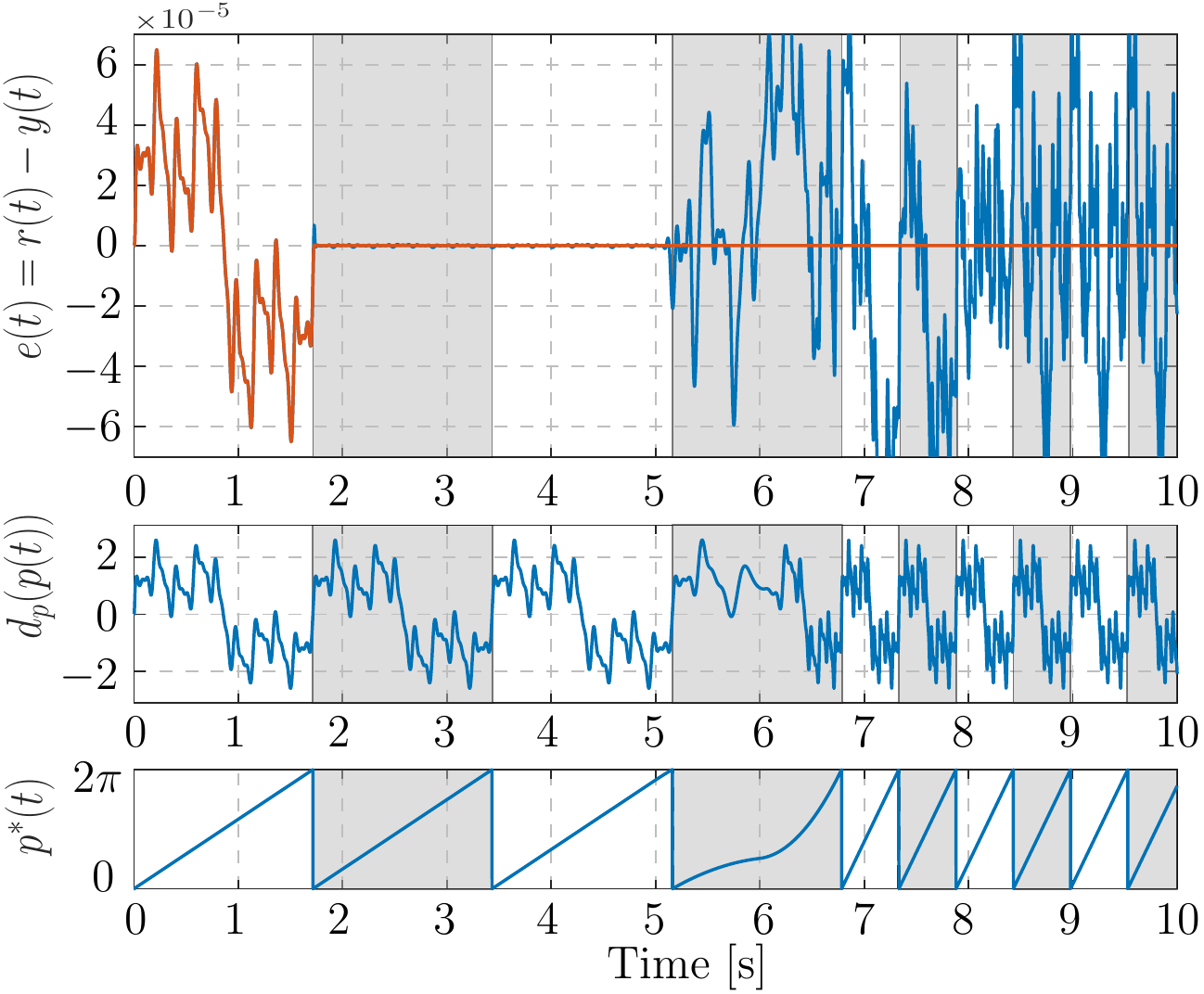}
		\caption[]{Top: Resulting positioning error for the traditional RC  (\tikz{\draw[mblue,line width=1.2pt]  (0,2.5pt)--(10pt,2.5pt); \draw  (0,0)}) and the spatial RC (\tikz{\draw[mred,line width=1.2pt]  (0,2.5pt)--(10pt,2.5pt); \draw  (0,0)}). Middle: corresponding disturbance signal. Bottom: positioning signal generating the disturbance. The individual periods are indicated with the white and gray areas.}
		\label{fig:RC_results_time}
	\end{centering}
\end{figure}

\subsection{Repetitive controller design}
The learning filter is designed according to Procedure \ref{proc:L_fitler}, i.e., equation \eqref{eq:learning_filter}. For traditional RC, the learning filter is designed as the inverse of the complementary sensitivity function, see \cite{Longman2010}. 

The positioning signal $p(t)$ that drives the disturbance is generated in a second equivalent loop, this allows to separate the effect of the reference and the disturbance. The velocity profile for the reference is equivalent to Fig. \ref{fig:Dist_var_velocity}, resulting in the position signal $p(t)$ in the bottom plot of Fig. \ref{fig:RC_results_time}. The disturbance mapping $d_p(p)$ is given by,
\begin{align*}
d_p(p) = 1.5\sin(p)+0.8\sin(3p)+ 0.6\sin(9p)+ \hdots \\ + 0.4\sin(18p)+ 0.2\sin(27p).
\end{align*}
and the time-domain disturbance is generated as in \eqref{eq:dist_gen}, i.e., at sample $k$ it is given by $d(k) = d_p(p(k))$, see Fig. \ref{fig:RC_results_time}. The period of the spatial RC is set to one revolution $p_\mathrm{per} = 2 \pi$ rad, and $\sigma_n = 10^{-6}$, $l = 0.1$, $\sigma_f = 1$. The memory loop size for traditional RC is chosen equal to the first disturbance period $N_\mathrm{conv} = 1717$.

\subsection{Simulation results}
\begin{figure}
	\begin{centering}
		\includegraphics[width=\linewidth]{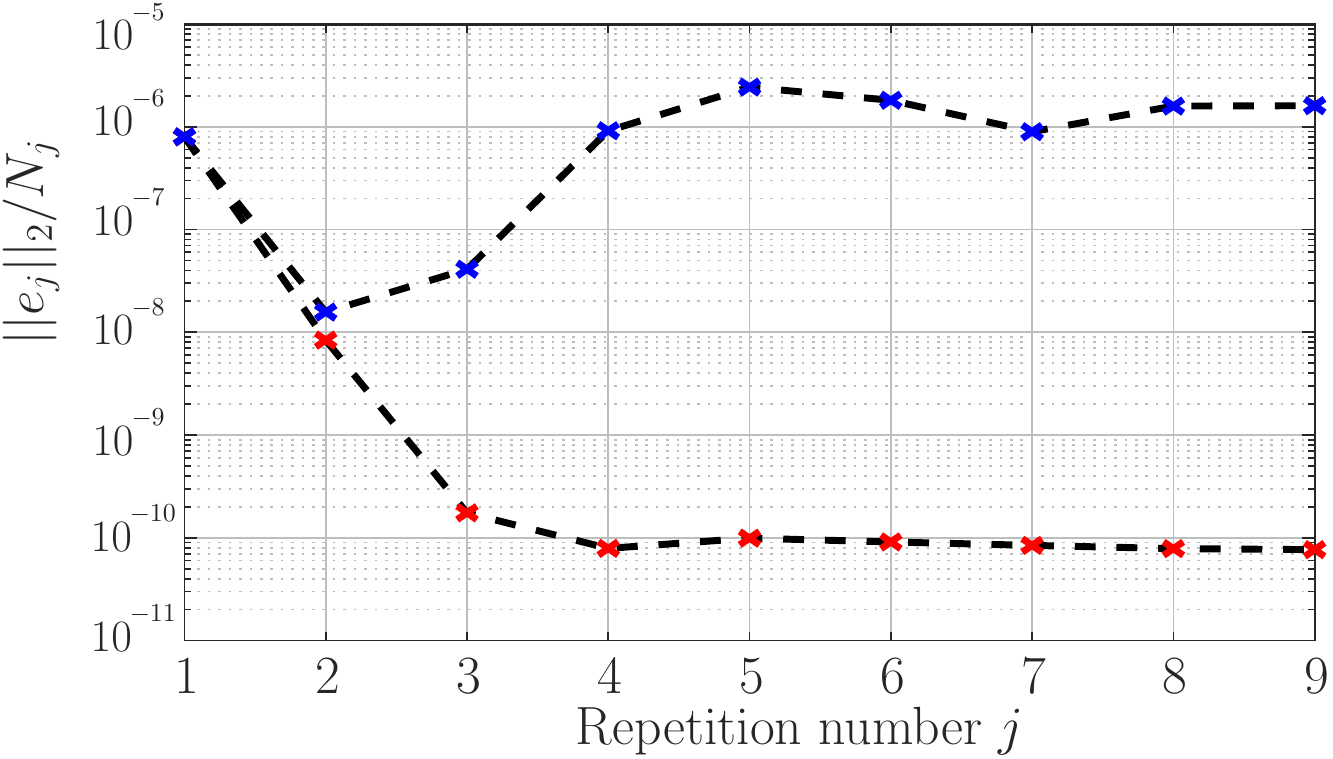}
		\caption[]{$2$-norm of the error normalized with the period length, for the traditional RC (\textcolor{mblue}{$\mathbf{\times}$}), and the spatial RC (\textcolor{mred}{$\mathbf{\times}$}) as function of the repetition number.}
		\label{fig:RC_results_var}
	\end{centering}
\end{figure}

A simulation is conducted using both traditional RC and spatial RC. The resulting errors and the applied disturbance are shown in Fig. \ref{fig:RC_results_time}, where the individual periods are indicated with white and gray areas. It can be seen that the disturbance is periodic during the first $3$ periods, after which a change in velocity results in a changing disturbance. To compare both methods, the $2$-norm of each period $j$ is normalized by the period length $N_j$, the result is depicted in Fig. \ref{fig:RC_results_var}. During the first period, the errors are equivalent since both RCs are not yet active. After one period it can be seen that both methods are able to significantly attenuate the disturbance leading to a small error. 
\begin{Remark}
Note that the traditional RC is not able to completely mitigate the disturbance. This occurs since the disturbance period time varies a little, due to its dependence on $p(t)$. Hence, it deviates slightly from the memory-loop size.
\end{Remark} After $3$ periods, the disturbance changes significantly leading to degraded performance in the traditional RC approach. It is not able to learn the disturbance since it is not compatible with the memory loop size. Hence, traditional RC requires re-tuning for changing velocities. The spatial RC is not affected by the changing velocity. Clearly it outperforms traditional RC, and continues to converge to a smaller error, see Fig. \ref{fig:RC_results_var}.

Finally, it is concluded that the developed spatial RC method outperforms traditional RC, by enabling attenuation of disturbances that are periodic in the spatial domain but potentially a-periodic in the time domain.

\section{Conclusion} \label{sec:conclusions}
Disturbances that are driven by position, may appear a-periodic in the time domain, due to velocity changes, yet have a periodic position-domain equivalent. In this paper, a new spatial RC approach is developed that attenuates these disturbances independent of velocity variations. A memory loop in the spatial domain based on a GP with suitable periodic kernel is developed. This effectively deals with non-equidistant observations inherent to the spatial-domain. A simulation example compares spatial RC with traditional RC and confirms the performance benefit, i.e., that it can deal with position dependent disturbances such as cogging, imbalances, etc., as encountered in many practical applications.

\appendix

\begin{thebibliography}{24}
	\providecommand{\natexlab}[1]{#1}
	\providecommand{\url}[1]{\texttt{#1}}
	\providecommand{\urlprefix}{URL }
	\expandafter\ifx\csname urlstyle\endcsname\relax
	\providecommand{\doi}[1]{doi:\discretionary{}{}{}#1}\else
	\providecommand{\doi}{doi:\discretionary{}{}{}\begingroup
		\urlstyle{rm}\Url}\fi
	
	\bibitem[{Ahn et~al.(2005)Ahn, Chen, and Dou}]{AhnCheDou2005}
	Ahn, H.S., Chen, Y., and Dou, H. (2005).
	\newblock State-periodic adaptive compensation of cogging and coulomb friction
	in permanent magnet linear motors.
	\newblock In \emph{Proceedings of the American Control Conference, 2005.},
	3036--3041. IEEE.
	
	\bibitem[{Blanken et~al.(2017)Blanken, Boeren, Bruijnen, and
		Oomen}]{BlankenBoeBruOom2017}
	Blanken, L., Boeren, F., Bruijnen, D., and Oomen, T. (2017).
	\newblock Batch-to-batch rational feedforward control: From iterative learning
	to identification approaches, with application to a wafer stage.
	\newblock \emph{IEEE/ASME Transactions on Mechatronics}, 22(2), 826--837.
	
	\bibitem[{Blanken and Oomen(2019)}]{BlankenOom2019}
	Blanken, L. and Oomen, T. (2019).
	\newblock Multivariable iterative learning control design procedures: From
	decentralized to centralized, illustrated on an industrial printer.
	\newblock \emph{IEEE Transactions on Control Systems Technology}.
	
	\bibitem[{Blanken and Oomen(2020)}]{BlankenOom2020_GP}
	Blanken, L. and Oomen, T. (2020).
	\newblock Kernel-based identification of non-causal systems with application to
	inverse model control.
	\newblock \emph{Automatica}, 114, 108830.
	
	\bibitem[{Chen and Chiu(2008)}]{ChenChi2008}
	Chen, C.L. and Chiu, G.T.C. (2008).
	\newblock Spatially periodic disturbance rejection with spatially sampled
	robust repetitive control.
	\newblock \emph{Journal of Dynamic Systems, Measurement, and Control}, 130(2),
	021002.
	
	\bibitem[{Chen and Yang(2007)}]{ChenYang2007}
	Chen, C.L. and Yang, Y.H. (2007).
	\newblock Adaptive repetitive control for uncertain variable-speed rotational
	motion systems subject to spatially periodic disturbances.
	\newblock In \emph{2007 American Control Conference}, 564--569. IEEE.
	
	\bibitem[{Chen et~al.(2012)Chen, Ohlsson, and Ljung}]{ChenOhlLju2012}
	Chen, T., Ohlsson, H., and Ljung, L. (2012).
	\newblock On the estimation of transfer functions, regularizations and gaussian
	processes revisited.
	\newblock \emph{Automatica}, 48(8), 1525--1535.
	
	\bibitem[{Francis and Wonham(1976)}]{FrancisWonham1976}
	Francis, B.A. and Wonham, W.M. (1976).
	\newblock The internal model principle of control theory.
	\newblock \emph{Automatica}, 12(5), 457--465.
	
	\bibitem[{Hara et~al.(1988)Hara, Yamamoto, Omata, and
		Nakano}]{HaraYamOmaNak1988}
	Hara, S., Yamamoto, Y., Omata, T., and Nakano, M. (1988).
	\newblock Repetitive control system: A new type servo system for periodic
	exogenous signals.
	\newblock \emph{IEEE Transactions on automatic control}, 33(7), 659--668.
	
	\bibitem[{Jidling et~al.(2018)Jidling, Hendriks, Wahlstr{\"o}m, Gregg,
		Sch{\"o}n, Wensrich, and Wills}]{JidlingHenWahGreSchoWenWill2018}
	Jidling, C., Hendriks, J., Wahlstr{\"o}m, N., Gregg, A., Sch{\"o}n, T.B.,
	Wensrich, C., and Wills, A. (2018).
	\newblock Probabilistic modelling and reconstruction of strain.
	\newblock \emph{Nuclear Instruments and Methods in Physics Research}, 436,
	141--155.
	
	\bibitem[{Li(2015)}]{Li2015}
	Li, P.Y. (2015).
	\newblock Prototype angle-domain repetitive control-affine parameterization
	approach.
	\newblock \emph{Journal of Dynamic Systems, Measurement, and Control}, 137(12).
	
	\bibitem[{Longman(2010)}]{Longman2010}
	Longman, R.W. (2010).
	\newblock On the theory and design of linear repetitive control systems.
	\newblock \emph{European Journal of Control}, 16(5), 447--496.
	
	\bibitem[{Mooren et~al.(2020)Mooren, Witvoet, A\c{c}an, Kooijman, and
		Oomen}]{MoorenWitAcaKooOom2020}
	Mooren, N., Witvoet, G., A\c{c}an, I., Kooijman, J., and Oomen, T. (2020).
	\newblock Suppressing position-dependent disturbances in repetitive control:
	With application to a substrate carrier system.
	\newblock In \emph{International Workshop on Advanced Motion Control (Accepted
		for publication)}.
	
	\bibitem[{Murphy(2012)}]{Murphy2012}
	Murphy, K.P. (2012).
	\newblock \emph{Machine learning: a probabilistic perspective}.
	\newblock MIT press.
	
	\bibitem[{Pillonetto et~al.(2014)Pillonetto, Dinuzzo, Chen, De~Nicolao, and
		Ljung}]{PillonettoDinCheNicLju2014}
	Pillonetto, G., Dinuzzo, F., Chen, T., De~Nicolao, G., and Ljung, L. (2014).
	\newblock Kernel methods in system identification, machine learning and
	function estimation: A survey.
	\newblock \emph{Automatica}, 50(3), 657--682.
	
	\bibitem[{Saathof et~al.(2019)Saathof, Crowcombe, Kuiper, van~der Valk,
		Pettazzi, de~Lange, Kerkhof, van Riel, de~Man, Truyens et~al.}]{Saathof2019}
	Saathof, R., Crowcombe, W., Kuiper, S., van~der Valk, N., Pettazzi, F.,
	de~Lange, D., Kerkhof, P., van Riel, M., de~Man, H., Truyens, N., et~al.
	(2019).
	\newblock Optical satellite communication space terminal technology at {TNO}.
	\newblock In \emph{International Conference on Space Optics}, volume 11180.
	International Society for Optics and Photonics.
	
	\bibitem[{Snoek et~al.(2012)Snoek, Larochelle, and Adams}]{SnoekLarAda2012}
	Snoek, J., Larochelle, H., and Adams, R.P. (2012).
	\newblock Practical bayesian optimization of machine learning algorithms.
	\newblock In \emph{Advances in neural information processing systems},
	2951--2959.
	
	\bibitem[{Steinbuch(2002)}]{Steinbuch2002}
	Steinbuch, M. (2002).
	\newblock Repetitive control for systems with uncertain period-time.
	\newblock \emph{Automatica}, 38(12), 2103--2109.
	
	\bibitem[{Tomizuka(1987)}]{Tomizuka1987}
	Tomizuka, M. (1987).
	\newblock Zero phase error tracking algorithm for digital control.
	\newblock \emph{Journal of Dynamic Systems, Measurement, and Control}, 109(1),
	65--68.
	
	\bibitem[{van~de Wijdeven and Bosgra(2010)}]{WijdevenBos2010}
	van~de Wijdeven, J. and Bosgra, O. (2010).
	\newblock Using basis functions in iterative learning control: analysis and
	design theory.
	\newblock \emph{Int. Journal of Control}, 83(4), 661--675.
	
	\bibitem[{van Zundert and Oomen(2017)}]{Zundert2017_Journal}
	van Zundert, J. and Oomen, T. (2017).
	\newblock On inversion-based approaches for feedforward and {ILC}.
	\newblock \emph{Mechatronics}, 50, 282 -- 291.
	
	\bibitem[{Wang et~al.(2009)Wang, Gao, and Doyle~III}]{WangGaoDoy2009}
	Wang, Y., Gao, F., and Doyle~III, F.J. (2009).
	\newblock Survey on iterative learning control, repetitive control, and
	run-to-run control.
	\newblock \emph{Journal of Process Control}, 19(10), 1589--1600.
	
	\bibitem[{Williams and Rasmussen(2006)}]{Rasmussen2006}
	Williams, C.K. and Rasmussen, C.E. (2006).
	\newblock \emph{Gaussian processes for machine learning}, volume~2.
	\newblock MIT Press Cambridge, MA.
	
	\bibitem[{Witvoet et~al.(2019)Witvoet, Peters, Kuiper, and
		Oomen}]{WitvoetPetKuiOom2019}
	Witvoet, G., Peters, J., Kuiper, S., and Oomen, T. (2019).
	\newblock Line-to-line repetitive control of a 6-{DoF} hexapod stage for
	overlay measurements using atomic force microspy.
	\newblock In \emph{Preceedings of American Control Conference (ACC),
		Philadelphia, PA, USA}.
	
\end{thebibliography}
\end{document}